\documentstyle[sprocl]{article}

\input{epsf.tex}
\def\Journal#1#2#3#4{{#1} {\bf #2}, #3 (#4)}

\def\PRD{{\em Phys. Rev.} D}

\def\be{\begin{equation}}
\def\ee{\end{equation}}
\def\bea{\begin{eqnarray}}
\def\eea{\end{eqnarray}}

\begin{document}

\title{Detection of Scalar Particles in Gravitational Waves from Resonant-Mass 
Detectors of Spherical Shape}

\author{E. Coccia}

\address{Universit\'a di Roma ''Tor Vergata"\\
Dipartimento di Fisica and I.N.F.N. sez. Roma II,\\ 
Via della Ricerca Scientifica, 00133 Roma, Italy}

\author{F. Fucito}

\address{I.N.F.N. sez. Roma II,\\ 
Via della Ricerca Scientifica, 00133 Roma, Italy}


\maketitle\abstracts{We report on some recent work, which points out the relevance
of future measurements of gravitational waves by resonant-mass detectors of spherical shape
for theories of gravity of non-Einstein type. }

\section{Introduction}
It is very likely that the next generation of 
detectors (both under contruction or in project) will be successfull
in the search for gravitational waves (GW).
Before giving the hard facts that motivates our optimism, we invite
the readers to look at Table 1.\\ \\ 
\begin{centering}
{\bf Table 1}: Possible sources of GW with their typical frequency. The last column
lists the way to detect such signals.
\end{centering}
\\
\begin{centering}
\footnotesize
\begin{tabular}{|c|c|c|l|}
\hline
Frequency (Hz) & Source & Detection Method \\
\hline
$10^{-16}$ & Primordial & Anisotropy of CBR\\
\hline
$10^{-9}$ & Primordial, Cosmic String & Timing of ms pulsars\\
\hline
 & Binary Starts & \\
$10^{-4}\div 10^{-1}$ & Supermassive BH & Laser interferometers in space\\
\hline
 & Binaries of NS or BH & \\
$10\div 10^{3}$ & Supernovae, Pulsars & Laser interferometers on earth\\
\hline
& Coalescence of NS or BH binaries & \\
$10^{3}$ &Supernovae, ms Pulsars & Resonant-mass detectors\\
\hline
\end{tabular}
\end{centering}\\
\vspace{0.2cm}\\
In the rest of this talk we will concentrate only on interferometers like LIGO
and VIRGO \cite{virgo},
which are under contruction, and resonant mass detectors as the already operating NAUTILUS
\cite{nautilus1,nautilus2} or the projected one of spherical shape \cite{virgo}.
The sensitivity of such instruments is plotted in Figure 2 from which
we infer that the new generation of detectors should be able to find GW's, 
given the already known sources (for a review see \cite{kt}) and given that the sensitivities
of the project will be attained in the final construction.
The idea that we want to convey to the reader is that this discovery can not only confirm 
Einstein's general relativity (GR), but can also lead us to probe new physics. How is this
possible? By measuring the spin content of the theory. Some of the most credited theories
for unifying all forces, predict, in fact, that in their gravitational sector the metric tensor
couples with scalars and higher spin tensors. Measuring scalar GW's seems to be the simplest of all
options. Out of all the detectors we have spoken of above, the resonant mass detector of spherical
shape is the one which is naturally privileged for this kind of measurement, given its geometrical
shape. We remark that also a certain number of interferometers, arranged in a geometrical 
precise pattern, could perform the same task \cite{harada}, having, moreover, the advantage of probing
a wider bandwidth in frequency with respect to the resonant mass detectors.
In the rest of this lecture we will discuss how the resonant mass detector of spherical shape
can probe the spin content of the theory of gravity and what is the strenght of the signal
given by scalar GW's with respect to metric tensor GW's in a binary system of stars.
\centerline{\vbox{\epsfysize=80mm \epsfbox{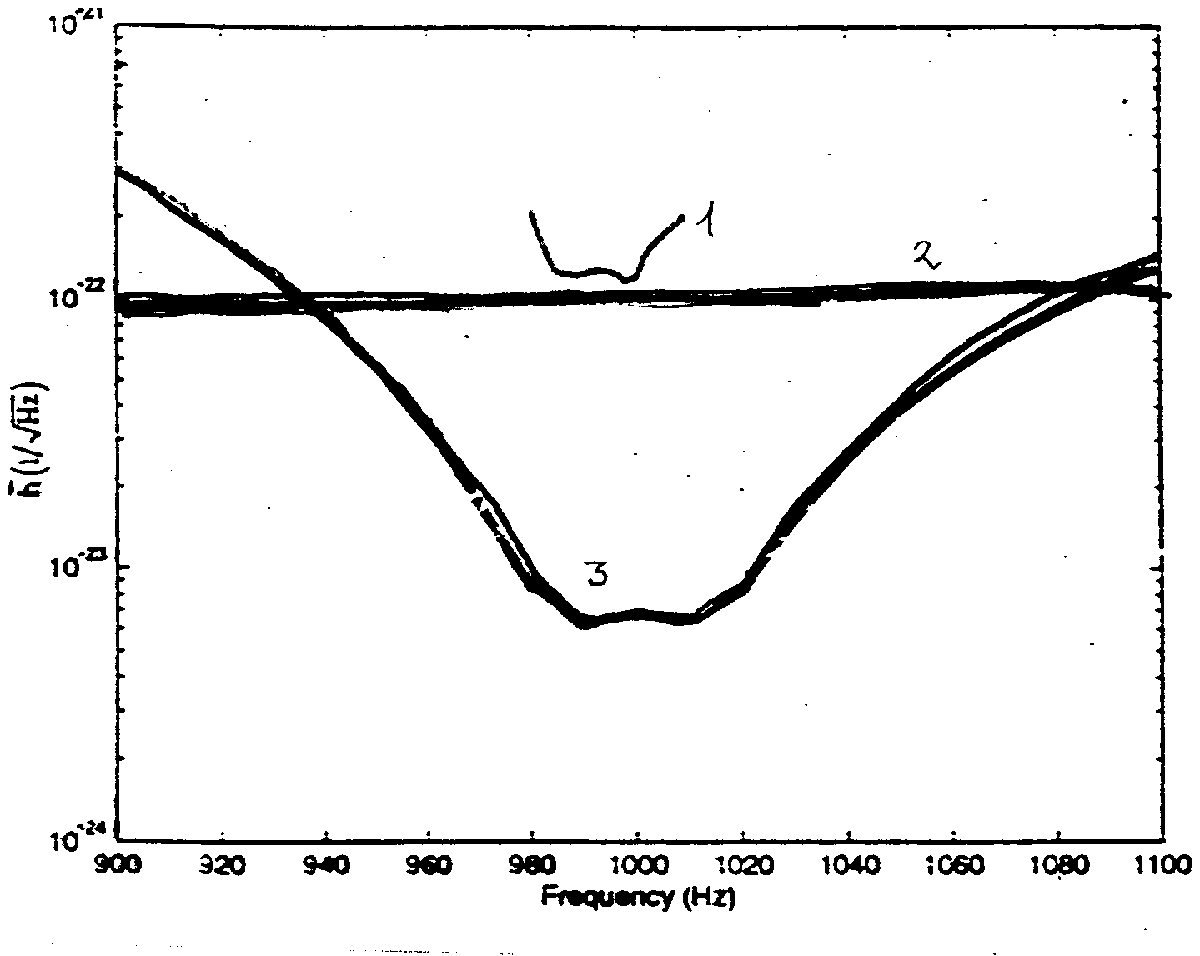}}}
\smallskip
\begin{centering}
{Figure 2: Sensitivity curves for interferometers (2) and resonant mass detectors of 
spherical (3) and cilindrical shape (1). }
\end{centering}
\section{A veto for metric theories}
The basic idea is very simple: the vibration frequencies (and their eigenfunctions) of 
a sphere can be divided into two sets called the spheroidal and toroidal modes.
When the sphere is made to interact with a gravitational theory of the Einstein type, the 
interaction with the toroidal modes is zero. Thus, monitoring the toroidal modes we can infer which
type of gravitational theory, we are dealing with \cite{lobo,bccff}.
We now give some details of this computation. All notations can be found in \cite{bccff}.
The equations of motion of the free vibrating sphere are 
\be
\rho \frac {\partial ^{2}u_{i}}{\partial  t^{2}}= \frac {\partial }{\partial
 x^{j}}(\delta_{ij}\lambda u_{ll}+2\mu u_{ij}) 
\label{esfera}
\ee
with the boundary condition:
\be
n_{j}\sigma _{ij}=0
\label{contorno}
\ee
The time-independent solutions of (\ref{esfera}) can be expressed as a sum of a longitudinal and 
two transverse vectors
\be 
\vec{u}(\vec{x})=C_{0}\vec{\nabla}\phi (\vec{x})+C_{1}
\vec{L}\chi (\vec{x})+C_{2}\vec\nabla\times\vec{L}\chi (\vec{x}) 
\label{usol}
\ee
where $C_{0}, C_{1}, C_{2}$ are dimensioned constants and $\vec{L}
\equiv\vec{x}\times
\vec{\nabla}$ is the angular momentum operator.
Regularity at $r=0$ restricts the scalar functions 
$\phi$ and $\chi$  
to be
expressed as $\phi(r,\theta,\varphi)\equiv j_{l}(qr)Y_{lm}
(\theta ,\varphi)$ and
$\chi(r,\theta,\varphi)\equiv j_{l}(kr)Y_{lm}(\theta ,\varphi)$. 
$Y_{lm}(\theta ,\varphi)$ are the spherical harmonics and $j_{l}$ the 
spherical Bessel functions.
Toroidal modes are obtained by setting $C_{0}=C_{2}=0$, and $C_{1}
\neq 0$, while spheroidal modes are obtained by 
setting $C_{1}=0$, $C_{0}\neq 0$ and $C_{2}\neq 0$. 
The displacement $\vec u$ of a point in the detector can be 
decomposed in normal modes as
\be
\vec u(\vec{x},t) = \sum_N A_N(t) \vec \psi_N (\vec x)
\label{expan}
\ee
where $N$ collectively denotes the set of quantum numbers 
identifying the mode.
The basic equation governing the response of the detector is 
\be
\ddot{A}_{N}(t) + \tau _{N}^{-1} \dot{A}_{N}(t) + 
\omega _{N}^{2}A_{N}(t) = 
f_{N}(t)
\label{forzate}
\ee
In terms of the so-called electric components of the Riemann tensor 
$E_{ij}\equiv R_{0i0j}$,
the driving force $f_N(t)$ is then given by 
\be
f_{N}(t) = - M^{-1} E_{ij}(t) \int \psi_{N}^{i*}(\vec{x}) x^{j}
\rho d^{3}x
\label{forza}
\ee
where $M$ is the sphere mass and we consider the density 
$\rho$ as a constant. When computed for the toroidal modes, (\ref{forza}) gives
zero.

\section{Binary systems}
In \cite{bcff} we computed the detector's signal-to-noise ratio defined as
\be 
SNR = {\Delta E_n \over \Delta E_{min}}
\label{sigtono}
\ee 
where $\Delta E_{min}$ is the minimum detectable energy innovation,
depending  on the detector's thermal and electronic noises.
The energy absorbed
by the detector's $n$-th mode is 
\be
\Delta E_n = \int_0^\infty \Phi(\omega) \sigma(\omega) d\omega \approx 2\pi 
\Phi(\omega_n) \Sigma_n
\label{formula}
\ee 
where $\Phi(\omega)$ is the incident GW energy flux per unit frequency.
Given the detector sensitivity and the differential cross section, computed in \cite{bbcfl},
it is possible to estimate the maximum distance at which the detector is able to probe the universe.
The computation for both laser interferometers and resonant mass detectors of spherical shape,
give a maximum distance, for scalar wave of the order of 100 kpar, to be compared with the
analogous
computation, for spin two which gave a result of the order of 1 Mpar.

\end{document}